\documentclass[preprint,authoryear]{elsarticle}
\usepackage[section] {placeins}
\usepackage{amsmath,amssymb,amstext,amscd,bm}
\journal{New Astronomy}
\def\astrobj#1{#1}

\begin{document}
\begin{frontmatter}

\title{Astrophysical and Structural Parameters of the Open Clusters {NGC~6866}, 
{NGC~7062}, and {NGC~2360}}

\author[kayseri]{Orhan  G\"une\c{s}\corref{cor1}\fnref{a}},
\author[istanbul]{Y\"uksel  Karata\c{s}\fnref{b}},
\author[PortoAlegre]{Charles Bonatto\fnref{c}}
\fntext[1]{E-mail: or.gunes@gmail.com}
\cortext[cor1]{Corresponding author}

\address[kayseri]{Department of Astronomy and Space Sciences, Faculty of Arts and Sciences, 
Erciyes University, Talas Yolu, 38039, Kayseri, Turkey}
\address[istanbul]{{\.I}stanbul University, Science Faculty, Department 
of Astronomy and Space Sciences, 34119 University, {\.I}stanbul, Turkey}
\address[PortoAlegre]{Universidade Federal do Rio Grande do Sul, Departamento 
de Astronomia, CP\,15051, RS, Porto Alegre 91501-970, Brazil}

\begin{abstract}

We derive astrophysical and structural parameters of the poorly studied open clusters
\astrobj{NGC~6866}, \astrobj{NGC~7062}, and \astrobj{NGC~2360} based on filtered 2MASS $(J, J-H)$ diagrams, and
stellar radial density profiles. The field star decontamination technique is utilised 
for selecting high-probability cluster members. The $E(B-V)$ reddening values of the three 
clusters derived from 2MASS JH${K_{s}}$ agree with those inferred from UBV and $uvby-\beta$ 
photometries. We find that the core mass function slopes are flatter than the halo's for the
three clusters. The large core and cluster radii of \astrobj{NGC~6866} and \astrobj{NGC~2360} indicate an expanded 
core, which may suggest the presence of stellar mass black-holes. \astrobj{NGC~2360} is located in the 
third quadrant ($\ell=229^{\circ}.80$), where Giant Molecular Clouds are scarce that,
together with its relatively large mass ($\sim1800~m_{\odot}$), might explain its longevity 
($\sim1.8$\,Gyr) in the Galaxy.
\end{abstract}

\end{frontmatter}

\section{Introduction\label{sec:s1}}

Reliable astrophysical and structural parameters of Galactic open clusters (OCs)
are an excellent source of information for interpreting their dynamics, and the 2MASS 
JH${K_{s}}$ photometric data base provides a unique opportunity to derive these kind 
of parameters in a uniform way. 

The stellar content of OCs undergoes internal and external changes related to, e.g. 
stellar evolution, mass segregation, and encounters with the disk and Giant Molecular 
Clouds (GMCs) \citep{lam06}, \citep{gie07}. Combined, these process produce a varying
degree of mass loss that may lead to the cluster dissolution into the field. For instance, 
the relative lack of old OCs in the solar vicinity can be partly explained by encounters 
with GMCs. The mass function slope $\chi$, relaxation time and evolutionary parameters
are required \citep{bon05, bon07a, bic06a} in order to be able to see mass segregation 
effects in the stellar content of OCs.
 
In this paper we focus on the above issues on three poorly studied OCs based on the 
improved field star decontamination procedure via 2MASS JH${K_{s}}$ photometry
\citep{bon07a, bon07b} and \citep{cam10}.

This paper is organised as follows. In Section~2 the 2MASS photometry and the field star 
decontamination algorithm (employed in the CMD analyses) are presented. The derivation of 
astrophysical and structural parameters, and a comparison with the literature, are given 
in Sections~3 to 5. Section~6 is dedicated to mass function properties of the clusters. 
Discussion of the results are in Section~7. Finally, in Section~8 a conclusion is given.

\section{2MASS JH${K_{s}}$ photometry and Field-star decontamination}

The JH${K_{s}}$ photometry of 2MASS\footnote{The Two Micron All Sky Survey, available at 
\textit{http://www.ipac.caltech.edu/2mass/releases/allsky/}} catalogue \cite{skr06} has 
been used to find the cluster members of the OCs \astrobj{NGC~6866}, \astrobj{NGC~7062}, and \astrobj{NGC~2360}. For 
this, VizieR\footnote{http://vizier.u-strasbg.fr/viz-bin/VizieR?-source=II/246.} was used 
to extract J, H, and ${K_{s}}$ 2MASS photometry for a large-area centred on these clusters.
The XDSS\footnote{Extracted from the Canadian Astronomy Data Centre (CADC), at 
\textit{http://www.cadcwww.dao.nrc.ca/}} R images of \astrobj{NGC~6866}, \astrobj{NGC~7062}, \astrobj{NGC~2360} 
are displayed in Fig.~1.

The technique used here for determining the cluster members of the three clusters is the 
field star decontamination procedure coupled to the 2MASS JH${K_{s}}$ photometry. It samples
photometric properties of the stars in a comparison field to (statistically) remove the field 
contamination from the cluster CMD. This technique has been  succesfully used by \cite{bon07a,bon07b}, 
\cite{bon08}, \cite{bic08a}, and \cite{cam09}, and applied to \astrobj{NGC~6866}, \astrobj{NGC~7062}, and \astrobj{NGC~2360}. 
First, the stellar surface density  $\sigma (stars\,\rm arcmin^{-2}$) of the raw data of these 
three clusters, computed for a mesh size of $3^{\prime}\times3^{\prime}$ and centred on the 
coordinates in Table 1, has been displayed in the top panels of Figs.~2$-$4. The isopleth surfaces 
of these clusters are presented in the bottom panels of Figs.~2$-$4. From the stellar radial density 
profile (hereafter RDP) of each cluster, built based on the JH${K_{s}}$ photometry extracted for the
WEBDA\footnote{obswww.univie.ac.at/WEBDA-Mermilliod \& Paunzen \ (2003)} coordinates in Table 1, the 
cluster radii are taken to be $R=19'.07$ (\astrobj{NGC~6866}), $R=7'.53$ (\astrobj{NGC~7062}) and $R=22'.63$ (\astrobj{NGC~2360}), 
respectively (see Col.~10 of Table~4). The stellar RDP is the projected number of stars per area around 
the cluster centre. To avoid oversampling near the centre and undersampling for large radii, the RDPs 
are built by counting stars in concentric rings of increasing width with distance to the centre. The 
number and width of rings are optimised so that the resulting RDPs have adequate spatial 
resolution with moderate $1\sigma$ Poisson errors. The residual background level of each 
RDP corresponds to the average number of CM-filtered stars measured in the comparison field.

Consequently, stars within these cluster radii are considered to be probable cluster members. 
As the stellar comparison field, a wide external ring $(R=15'-20')$ has been considered to eliminate 
field stars. As \cite{cam10} noted, RDPs of OCs built based on the WEBDA coordinates usually show a dip in the 
inner RDP region. For this reason, new central coordinates of these clusters have been searched to maximise the star 
counts at the innermost RDP bin. Then, the 2MASS photometry was extracted again, but now centered on the 
optimised cluster coordinates. The optimised central coordinates are displayed in the panels of Fig.~1 as small 
circles, and given in the right section of Table 1. To have the intrinsic CMD morphology of the clusters, 
the statistical field star decontamination procedure of \cite{bon07a} is used. This procedure is 
based on the relative number densities of stars in a cluster region and offset field. It divides the full 
range of magnitudes and colours of a CMD into the cell dimensions of $\Delta{J}=1.0$, and  
$\Delta(J-H)={\Delta(J-K_{s})}=0.15$. 
These dimensions are adequate to allow sufficient star counts in individual cells and preserve the intrinsic
morphology of the evolutionary sequences. As shown in \cite{bon07a}, the field star decontamination 
procedure with 2MASS JH${K_{s}}$ is efficient in isolating stars with a high probability of being cluster 
members. Details on the algorithm can be found in \cite{bon07a, bon07b} and  \cite{bon09a, bon09b, bon09c}, 
and \cite{cam10}. 

\section{Astrophysical parameters}

Having applied the field star decontamination technique to eliminate field stars, the decontaminated 
$(J, J-H)$ CMDs of \astrobj{NGC~6866}, \astrobj{NGC~7062}, and \astrobj{NGC~2360} are plotted in Figs.~5(a)$-$(c). The shaded area 
in the figures are the colour-magnitude filters, which follow the distribution of the decontaminated
star sequences in the CMDs. These filters are wide enough to accommodate 
the colour distributions of main sequence and evolved stars of the clusters, allowing 1 $\sigma$
photometric uncertainties. 
The solid lines in Figs.~5(a)$-$(c) represent the fitted 0.8, 1 and 
1.8 Gyr Padova isochrones (Girardi et al. 2002, hereafter G02) for $Z= +0.019$ (solar) abundance. 
As can be seen from Figs.~5(a)$-$(c), the G02 isochrones fit well the main sequence (MS), 
turn$-$off (TO) and RC/RG regions on the CMDs of the clusters. 
Due to the presence of binaries, the G02 isochrones have been shifted to the left of the 
main sequence in Figs.~5a$-$c \citep[see the Sect.~5]{bon08}.

The astrophysical parameters (reddening, distance modulus, distance and age) of three clusters 
have been derived based on the solar metallicity isochrones with the ages 0.8, 1 and 1.8 Gyr G02 
isochrones fitted to the decontaminated CMDs. We find the reddenings $E(J-H)$ = 0.06$\pm$0.02 for 
\astrobj{NGC~6866}, $E(J-H)$ = 0.10$\pm$0.02 for \astrobj{NGC~7062}, $E(J-H)$ = 0.02$\pm$0.01 for \astrobj{NGC~2360}, respectively,
from the CMD diagrams of Figs.~5(a)$-$(c). These reddenings in $E(J-H)$ are converted to 
$E(B-V)$ by considering the relations $A_{J}/{A_{V}}=0.276$, $A_{H}/{A_{V}}=0.176$, 
$A_{K_{s}}/{A_{V}}=0.118$, $A_{J}=2.76\times{E(J-H)}$, and $E(J-H)=0.33\times{E(B-V)}$ 
\citep{dut02}, assuming a constant total-to-selective absorption ratio $R_{V}=3.1$.
The resulting $E(B-V)$ values are given in Col.~3 of Table 2. Following the same procedure,
the distance moduli of these clusters have been derived and listed in Col.~4 of Table 2. The 
estimated heliocentric $d~(kpc)$ and Galactocentric $R_{GC}$ (kpc) distances are given in Cols.~5$-$6, 
respectively. When estimating the $R_{GC}$ distances, the value $R_{\odot}=7.2\pm0.3$ kpc of 
\cite{bic06b} is considered. 

The reddening values of \astrobj{NGC~6866}, \astrobj{NGC~7062}, and \astrobj{NGC~2360} have been compared to those of 
the dust maps of \citet[hereafter SFD]{sch98}. 
These are based on the COBE/DIRBE and IRAS/ISSA maps, and take into account
the dust absorption all the way to infinity.  SFD dust maps give $E(B-V)_{\infty}$ 
as 0.72 for \astrobj{NGC~6866}, 1.15 for \astrobj{NGC~7062}, 0.80 for \astrobj{NGC~2360}.  
The distances in Col.~5 of Table 2 is used for the reduced reddening.
By following the method of \cite{boni}, the reduced final reddenings are 
$E(B-V)_{A} =$ 0.42,~0.42, and 0.18 for \astrobj{NGC~6866}, \astrobj{NGC~7062}, and \astrobj{NGC~2360}, respectively. 
Within the uncertainties the reddening value of $0.32\pm0.06$ of 2MASS JH${K_{s}}$ photometry for \astrobj{NGC~7062}  is 
close to the value of $0.42$ obtained from the dust maps of SFD.  
Our reddening values of 0.19 for \astrobj{NGC~6866} and 0.06 for \astrobj{NGC~2360} are 
lower than $E(B-V)$=0.42 and 0.18 of the dust maps of SFD, respectively.

\section{Comparison with the literature}

A comparison of the present astrophysical parameters with those in the literature is 
given in Table 3. Our (2MASS) value of $E(B-V)=0.19\pm0.06$ for \astrobj{NGC~6866} agrees with
the $(0.17,~UBV)$ of \cite{hoa61}; the same applies to our $E(B-V)=0.06\pm0.03$ of 
\astrobj{NGC~2360} with respect to the $(0.07,~UBV)$ of \cite{egg68}. However, our reddening 
value of $E(B-V)=0.32\pm0.06$ of \astrobj{NGC~7062} is relatively far from the $(0.46,~ubvy-\beta)$ 
of \cite{pen90}, which might be partly explained by the very different metallicities 
used in both studies. 

As can seen in Cols.~5$-$6 of Table 3, within the uncertainties the literature distance 
moduli and distances of \astrobj{NGC~6866} and \astrobj{NGC~2360} are relatively close to the values of this 
paper. The age of \astrobj{NGC~6866} (from the 2MASS JH${K_{s}}$ photometry) of this paper is 
older than 0.38 Gyr value of \cite{hoa61}. Although the reddenings and heavy element 
abundances between this paper and \cite{hoa61} are almost the same, this difference in ages results from
the adopted isochrones. The older study of \cite{hoa61} considers the observational ZAMS. 
This paper considers the G02 isochrone, which contains the new input physics.
Since \cite{egg68} did not give an age value for \astrobj{NGC~2360}, no any comparison was done. 
 
The distance modulus and distance of \astrobj{NGC~7062} are fainter and larger as compared to 
the result of \cite{pen90}. Our age value of 1.0 Gyr, derived from 2MASS JH${K_{s}}$ 
photometry is quite older than the 0.28 Gyr given by \cite{pen90}. These differences stem 
from the metal and heavy element abundance approximations between 
this paper and \cite{pen90}. \cite{pen90} measure the abundances of \astrobj{NGC~7062} as 
$([Fe/H],~Z)=(-0.35,~+0.003)$ from $ubvy-\beta$ photometry, and consider the 
isochrone of \cite{van85} for $Z=+0.003$, 
whereas we adopt the solar metallicity G02 isochrones.

\section{Structural parameters}

Structural parameters of \astrobj{NGC~6866}, \astrobj{NGC~7062}, and \astrobj{NGC~2360} are derived by means of
the stellar radial density profile (RDP), which is explained in Section~2.
Usually, the RDPs of star clusters can be described by an analytical profile, like the 
empirical, single mass, modified isothermal spheres of \cite{kin66} and \cite{wil75}, 
and the power law with a core of \cite{e87}. These functions are characterised by different 
sets of parameters that are related to the cluster structure. 
We adopt the two-parameter function $\sigma(R) = \sigma_{bg} + 
\sigma_0/(1+(R/R_c)^2)$, where $\sigma_{bg}$ is the residual background density, $\sigma_0$ 
the central density of stars, and $R_{core}$ the core radius. Applied to star counts,
this function is similar to that used by \cite{kin62} to describe the surface brightness 
profiles in the central parts of globular clusters. 
The RDPs  of the clusters \astrobj{NGC~6866}, \astrobj{NGC~7062}, and \astrobj{NGC~2360}, fitted with King profiles are
shown in Figs.~6(a)$-$(c). In Fig.6 the solid line shows the best fit King profile,  
horizontal red bar denotes the stellar background level measured 
in the comparison field, and the $1\sigma$ King fit uncertainty is shown by the shaded domain.
The cluster radius ($R_{RDP}$) is also obtained from the measuring the distance from 
the cluster centre where the RDP and residual 
background are statistically indistinguishable \citep{bon07a}. The $R_{RDP}$ can 
be taken as an observational truncation radius, whose value depends both on the radial 
distribution of member stars and the field density.
These structural parameters and their meanings are listed in Table 4.  

\section{Mass and Mass functions}

The stellar masses stored in \astrobj{NGC~6866}, \astrobj{NGC~7062}, and \astrobj{NGC~2360} have been determined by means 
of their mass functions (MFs), built for the observed MS mass range according to \citet{bic06a}. 
By following the algorithm, which is basically defined by \citet{bon05}, 
luminosity functions from the decontaminated $(J, J-H)$ diagrams of the three clusters
have been transformed into MFs through the corresponding mass-luminosity relations derived
from Padova isochrones for the ages in Col.~2 of Table 2. 
The relations of $\phi(m)(stars ~m_\odot^{-1})$ versus $m_\odot$ of \astrobj{NGC~6866}, \astrobj{NGC~7062}, and
\astrobj{NGC~2360} are shown in Figs.~7$-$9 for different cluster regions. 
The mass ranges shown in Figs.~7$-$9 are $m=1.13-2.23~m_{\odot}$ for \astrobj{NGC~6866}, $m=1.08-1.93~m_{\odot}$ 
for \astrobj{NGC~7062}, and $m=0.83-1.63~m_{\odot}$ for \astrobj{NGC~2360}, respectively.

The MS MFs in Figs.~7$-$9 are fitted with the function $\phi(m)\propto{m}^{-(1+\chi)}$, and
the MF slopes ($\chi$) have been determined for the different regions, 
which are indicated in Figs.~7$-$9 and Col.~1 of Table 5.
Details of this approach are given in Table 5, where we also show the number and 
mass of the evolved stars. The lower MS is not accessed on the $(J, J-H)$ diagrams 
of the three clusters, but we assume that the low-mass content is still present, 
and use Kroupa's MF\footnote{$\chi=0.3\pm0.5$ \citep{kr01}
for $0.08<m_\odot<0.5$, $\chi=1.3\pm0.3$ for $0.5<m_\odot<1.0$, and $\chi=1.3\pm0.7$ 
for $1.0<m_\odot$.} to estimate the total stellar mass, down to the H-burning mass limit 
($0.08\,m_\odot$). The results (number of stars, MS and evolved star stellar contents, MF 
slope ($\chi$), and mass extrapolated to 0.08~$m_\odot$) for each cluster region are given 
in Table 5. 

The MF slopes of \astrobj{NGC~6866} (Fig.~7) are quite flat ($\chi=-1.37\pm0.54$) in the core 
and very steep ($\chi=+2.23\pm0.49$) in the region of $r=[1.97,~9.15]$ pc. Such a
MF slope steeping from the core to the outskirts of a cluster implies that core
low-mass stars are being transferred to the cluster's halo, 
while massive stars accumulate in the core, due to mass segregation.
The overall MF slope ($\chi=+1.32\pm0.46$) for $r=[0,~9.15]$ pc is quite close to the 
standard  Salpeter IMF value \citep{sal55} of $\chi=+1.35$.

The MF slope is also flat ($\chi=-0.66\pm0.60$) in the core of \astrobj{NGC~7062} cluster,  
whereas MF slope is very steep with $\chi=+2.28\pm0.44$ in the region of $r=[0.66,~3.59]$ pc.
The overall MF slope is also very steep with $\chi=+2.16\pm0.33$ for $r=[0,~3.59]$ pc, which is 
different from Salpeter's value. The numbers of low mass and high mass stars 
in the core of \astrobj{NGC~7062} are almost equivalent. 
As seen from Fig.~8, the variation in $\chi$ is quite large from the core 
to the halo of the cluster \astrobj{NGC~7062}, due to the large scale mass segregation. 

For the cluster \astrobj{NGC~2360}, the MF slope is quite flat ($\chi=-0.14\pm0.60$) 
in the core, increasing to $\chi=+1.07\pm0.48$ in the region 
$r=[1.13,~6.85]$ pc. The overall MF slope is $\chi=+0.67\pm0.36$ for $r=[0,~6.85]$ pc, 
which is  different from Salpeter's value. Flat core and overall MF slopes 
of this cluster imply mild mass segregation effects.

\section{Discussion}

The relaxation time $t_{rlx}$ is the characteristic time-scale for a cluster to reach 
some level of energy equipartition \citep{bin98} and, as discussed 
in \cite{bon05}, \cite{bon06}, \cite{bon07a}, the evolutionary 
parameter ($\tau = Age/t_{rlx}$) appears to be a good indicator of dynamical state. Following
\cite{bon06}, we parameterize $t_{rlx}$ as $t_{rlx}\approx0.04\left(\frac{N}{ln N}
\right)\left(\frac{R}{1pc}\right)$, where N is the number of stars located inside
the region of radius R. 
The uncertainties in the evolutionary parameters ($\tau$) of the three clusters have been 
estimated by propagating the errors in Age (Table 2), Radii (Table 4) and N (Table 5) into 
$t_{rlx}$ and $\tau$. Although the errors in cluster radii and age (Tables 2 and 4) are at a similar 
($\sim10-20\%$) level for the three clusters, the uncertainties in $R_{core}$ and especially in the
number of stars N are much larger, particularly for  \astrobj{NGC~7062} and \astrobj{NGC~2360}. When propagated,
the latter two errors produce a large uncertainty in $t_{rlx}$ (Table 6) and, consequently,
a large uncertainty in the evolutionary parameter. In this sense, both $t_{rlx}$ and $\tau$
should be taken simply as an order of magnitude estimate.

As emphasized by \cite{bon06}, the two parameters present a signature that low-mass stars 
originally in the core are transferred to the cluster's outskirts, 
while massive stars sink in the core, which is related to mass-segregation.
\citet{bon05,bon06} discuss that the significant flattening in core and overall MFs due 
to dynamical effects such as mass segregation is expected to occur for 
$\tau_{\rm core}\geq100$ and $\tau_{\rm overall}\geq7$, respectively. 

For a typical $\sigma_{v}\approx3\,km s^{-1}$ \citep{bin98}, the relaxation times 
$[t_{rlx}(overall),~t_{rlx}(core)]$ and the evolutionary parameters 
$[\tau_{overall},~\tau_{\rm core}]$ for  \astrobj{NGC~6866}, \astrobj{NGC~7062}, and \astrobj{NGC~2360} clusters 
are given in Table 6. 

The very steep core MF of \astrobj{NGC~6866} ($\chi=-1.37$) is due to the relaxation time and 
the evolutionary parameter of $[t_{rlx}(core),~\tau_{core}]$=$[1.77~Myr,~453]$. 
The very steep halo and rather steep overall 
MFs $(\chi=+2.23,\chi=+1.32)$ present signs of large-scale mass segregation in 
the core/halo region, due to the $[t_{rlx}(overall),~\tau_{overall}]$=$[210~Myr,~3.80]$. 

For \astrobj{NGC~7062}, the flat slope ($\chi=-0.66$) in the core is not coincident 
with the values of $[t_{rlx}(core),~\tau_{core}]$=$[0.61~Myr,~1640]$. 
$t_{rlx}(core)$ of this cluster is lower than the value of 7, 
given by \cite{bon06} and \cite{bic06a}, and quite short to occur a flat core 
because of mass segregation. A very flat core would have been expected 
for this cluster, due to large value of $\tau_{core}$. 
The very steep halo and overall MFs ~$(\chi=+2.28,\chi=+2.16)$ are in 
reasonable concordance with the relaxation times 
$[t_{rlx}(overall),~\tau_{overall} ]$=$[77.80~Myr,~12.80]$. 
The evolutionary parameters for this cluster suggest an advanced dynamical state.

With the $[t_{rlx}(core),~\tau_{core}]$=$[3.17~Myr,~569]$, \astrobj{NGC~2360} has
a flat core MF ($\chi=-0.14$). The rather steep halo and flat overall MFs~$(\chi=+1.07,\chi=+0.67)$ 
of this cluster present signs of mild mass segregation in the core/halo region, 
as a result from the $[t_{rlx}(overall),~\tau_{overall}]$=$[153~Myr,~11.80]$.

As seen from the Galactocentric distance values in Col.~6 of Table~2 and Fig.~10, 
since these clusters are located at the Solar circle, tidal disruption due to both 
the Galactic field and GMCs seems to be efficient.

The cluster and core radii $(R_{RDP},~R_{core})$ of the three clusters (Cols.~5$-$6 of Table 4),
confirm that the larger clusters tend to have larger cores, as shown by \citet[Fig.13(a)]{cam10}.
The cluster and core radii $(R_{RDP},~R_{core})$=$(3.59,~0.66)$ pc of \astrobj{NGC~7062}, 
which is quite small, imply that it may have shrunk in size with time. Thus, for this cluster, 
we see the dynamical effects of both mass segregation and core collapse.
The overall mass $m=1560~m_\odot$ of \astrobj{NGC~7062} also indicates that it is a survivor,
despite these dynamical processes and the location of $R_{GC}=7.30$ kpc.

\astrobj{NGC~6866} and \astrobj{NGC~2360} have large cluster and core radii (Cols.5$-$6 of Table 4), 
a fact that may be partly associated with 
large-scale mass segregation and/or related to initial conditions. Alternatively, 
cluster heating due to stellar mass black-holes that are scattered towards the halo 
or ejected from the cluster may also explain part of the expansion \citep{mac07}. 
Negative values of $\chi$ of the cores also show that 
the massive stars tend to be concentrated in the cores of the two clusters.

The relation between core and cluster radii, the dependence of the cluster radii on 
the Galactocentric distance, and the relation between the core and cluster radii with 
the age of the three clusters 
are presented in Figs.11(a)$-$(d), together with the data of \cite{cam09}.
\astrobj{NGC~7062} and \astrobj{NGC~6866} are almost near the bifurcation age 
of $\approx1$ Gyr, given by \cite{cam09}. \astrobj{NGC~2360} is slightly away from this bifurcation limit. 
The relation of the core and cluster radii with age is presented in Figs.11(c)$-$(d).
This relation is related to the cluster survival/dissociation rates, as emphasized by \cite{cam09}.
As is seen from panel~(b), the three clusters are inside the standard IAU value $R_{GC}=8.5$ kpc, 
and quite close to the $R_{\odot}=7.2\pm0.3$ kpc of \cite{bic06b}.

\astrobj{NGC~7062} has small cluster and core radii, $(R_{RDP},~R_{core})=(3.59,~0.66)$ pc,
and is located in the shaded region, plotted from the data (open circles) of  \cite{cam09} (panel a). 
The flat core MF slope ($\chi=-0.66\pm0.60$) and very steep overall MF slope ($\chi=+2.28\pm0.44$) of this cluster
show large scale mass segregation. \astrobj{NGC~7062} seems to be at stage of core shrinkage because of dynamical relaxation.
A nebulae near this cluster can cause the small core and cluster sizes, as a result of destruction process.
Alternatively, the small core and cluster sizes may be primordial, which may be related to the high molecular 
gas density in Galactic center directions \citep{van91,cam10}.

\astrobj{NGC~2360} is located in the third quadrant ($\ell=229^{\circ}.80$), where GMCs are scarce. 
Its longevity is probably related to the 
relatively large mass, $1800~m_{\odot}$, large cluster radii and location in the Galaxy.
Also, flat core and overall MF slopes of \astrobj{NGC~2360} imply mild mass segregation effects.

\astrobj{NGC~6866} is located at $\ell=79^{\circ}.56$, as is seen in Table 1. 
Due to the presence of GMCs, tidal effects from disk and Bulge crossings 
as external processes in this direction, this cluster is a survivor with the values 
of $1800~m_{\odot}$ and $R_{RDP}=9.15$ pc. Moreover, the very steep core slope 
$\chi=-1.37$ and quite steep overall slope $\chi=+1.32$ of \astrobj{NGC~6866} indicate that 
the mass segregation effects are quite efficient.

\section{Conclusions}

Our main conclusions are summarized as follows:

\begin{enumerate}

\item The astrophysical and structural parameteres of the poorly studied clusters , \astrobj{NGC~6866}, 
\astrobj{NGC~7062}, and \astrobj{NGC~2360} have been derived from the filtered 2MASS $(J, J-H)$ CMDs, and
the stellar RDPs. The field star decontamination technique is utilised 
for separating the cluster members. 

\item The astrophysical parameters (Age,~E(B-V),~d) are (0.80~Gyr,~0.19,~1.65 kpc) for \astrobj{NGC~6866},
(1.0~Gyr,~0.32,~1.64 kpc) for \astrobj{NGC~7062}, (1.80~Gyr,~0.06,~1.04 kpc) for \astrobj{NGC~2360}, respectively. 
These clusters are inside the Solar circle. 
The reduced final reddenings from the dust maps of SFD are 
$E(B$--$V)_{A}=$ 0.42,~0.42, and 0.18 for \astrobj{NGC~6866}, \astrobj{NGC~7062}, \astrobj{NGC~2360}, respectively. 
The reddening value of $0.32\pm0.06$ of 2MASS JH${K_{s}}$ photometry for \astrobj{NGC~7062} is 
close to the  value of $0.42$ obtained from the dust maps of SFD.  
Our reddening values of $E(B-V)=$0.19 and 0.06 for \astrobj{NGC~6866} and \astrobj{NGC~2360}, respectively 
are lower than $E(B-V)=$0.42 and 0.18 of the dust maps of SFD. 
However, the SFD values resulted from line-of-sight integral throughout 
the Milky Way and with low spatial resolution, it is quite a normal thing 
to have different reddening values for these relatively close ($\sim1$~kpc) star clusters.

\item The reddening values, 0.19 and 0.06 for \astrobj{NGC~6866} and \astrobj{NGC~2360} clusters, 
derived by  2MASS JH${K_{s}}$ photometry are in good agreement with the values of 
$(0.17,~UBV)$ of \cite{hoa61} and $(0.07,~UBV)$ of  \cite{egg68}. 
However, our reddening value of $E(B-V)=0.32\pm0.06$ of \astrobj{NGC~7062} is relatively far 
from the $(0.46,~ubvy-\beta)$ of \cite{pen90}, which might be partly explained by 
the very different metallicities used in both studies. 
Within the uncertainties, distance moduli and distances of \astrobj{NGC~6866} and \astrobj{NGC~2360} agree
with the values of this paper. The distance modulus and distance of \astrobj{NGC~7062} are lower 
than the one of \cite{pen90}. 
The 0.80 Gyr age value of \astrobj{NGC~6866}, derived from 2MASS JH${K_{s}}$ photometry, is 
older than 0.38 Gyr value of \cite{hoa61}. Although the reddenings and heavy element 
abundances between this paper and \cite{hoa61} are almost the same, this difference 
in ages are due to the usage of distinct isochrones which corresponds 
to differing internal physics of the isochrones and photometric systems.
The 1.0 Gyr age value of \astrobj{NGC~7062}, derived from 2MASS JH${K_{s}}$ 
photometry, is quite older than the 0.28 Gyr given by \cite{pen90}. 
These differences stem from the metal and heavy element abundance assumptions between 
this paper and \cite{pen90}.

\item \astrobj{NGC~6866} has MF slopes that are quite flat ($\chi=-1.37$) in the core 
and very steep ($\chi=+1.32$) in the region of $r=[0,~9.15]$ pc. This suggests 
that low-mass stars in the core are being transferred to the cluster's outskirts, 
while massive stars accumulate in the core. \astrobj{NGC~6866} is in the direction of the 
Galactic centre, and due to the presence of GMCs, and the dissolution effects 
associated with tidal effects from disk and Bulge crossings, this cluster is a 
survivor with the values of $1800~m_{\odot}$ and $R_{RDP}=9.15$ pc. Moreover, 
this cluster has large core and cluster radii, and shows an expanded core possibly 
due to the presence of a stellar mass black-hole.

\item The flat core MF slope ($\chi=-0.66$) and the steep overall MF slope ($\chi=+2.16$) indicate that 
the variation in $\chi$ is quite large from the core to the outskirts of the cluster \astrobj{NGC~7062}, 
due to the large scale mass segregation. 
With small cluster and core radii, $(R_{RDP},~R_{core})=(3.59,~0.66)$ pc,
this cluster seems to be at stage of core shrinkage because of dynamical relaxation.
A nebulae near this cluster may be responsible for the small core and cluster sizes, 
as a result of destruction process.
The small core and cluster sizes may be primordial, probably related to the high molecular 
gas density in Galactic center directions \citep{van91, cam10}.

\item The large core and cluster radii of \astrobj{NGC~2360} indicate an expanded core,
which may suggest the presence of a stellar mass black-hole.
\astrobj{NGC~2360} is located in the third quadrant ($\ell=229^{\circ}.80$), 
where GMCs are rare that, together with the large mass and cluster radius,
might explain its longevity ($\sim1.8$\,Gyr).
Also, flat core ($\chi=-0.14$) and overall MF ($\chi=+1.07$) slopes 
of this cluster imply mild mass segregation effects.

\end{enumerate}

\clearpage

\begin{figure}
\centering
\includegraphics*[width = 6cm, height = 12cm]{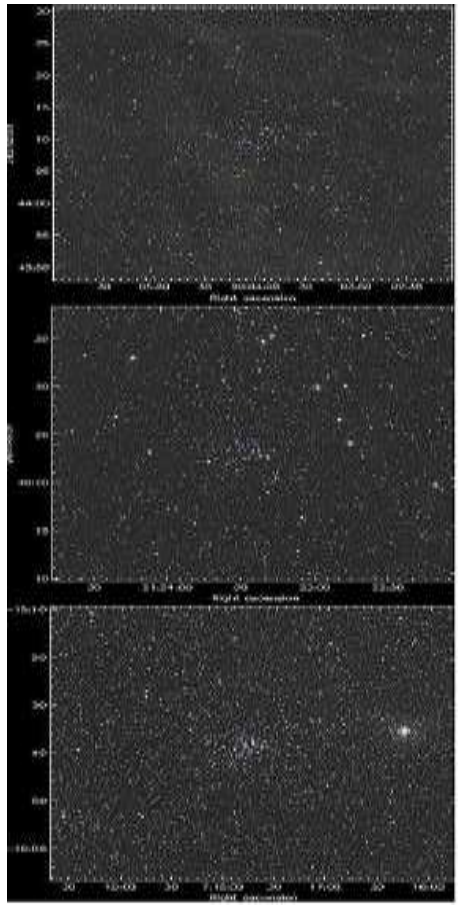}
\caption{The images of \astrobj{NGC~6866} DSS-R $30\,^\prime$x $30\,^\prime$, \astrobj{NGC~7062} DSS-R $20\,^\prime$x $20\,^\prime$ 
and \astrobj{NGC~2360} DSS-R $40\,^\prime$x $40\,^\prime$,
from top to bottom.}
\end{figure}

\begin{figure}
\centering
\includegraphics*[width = 10cm, height = 11cm]{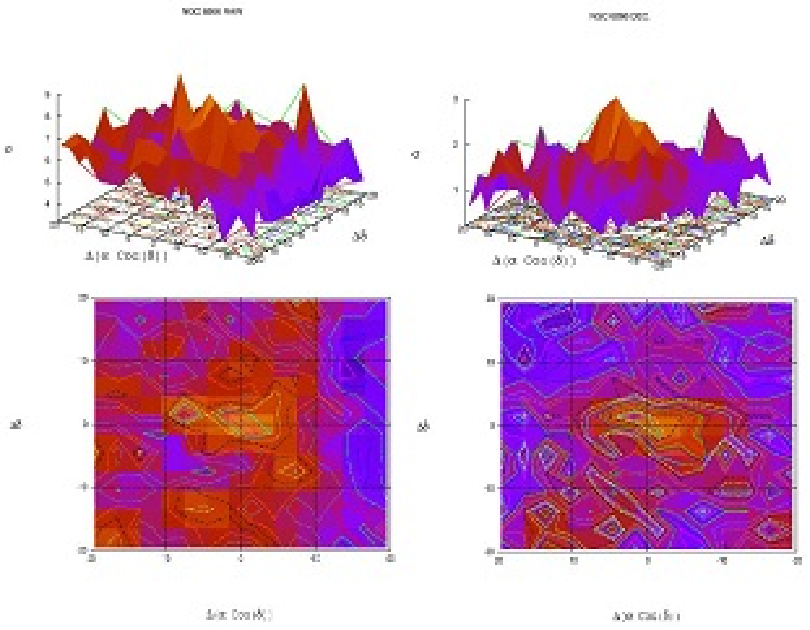}
\caption{For observed (raw) photometry, top panels: stellar surface$-$density 
$\sigma (stars\,\rm arcmin^{-2}$) of \astrobj{NGC~6866}, computed for a mesh size of 
$3^\prime\times3^\prime$, centred on the coordinates in Table 1. 
Bottom panels : The corresponding isopleth surfaces. 
Left panel: observed (raw) photometry. Right panel: decontaminated photometry.}
\end{figure}

\begin{figure}
\centering
\includegraphics*[width = 10cm, height = 10cm]{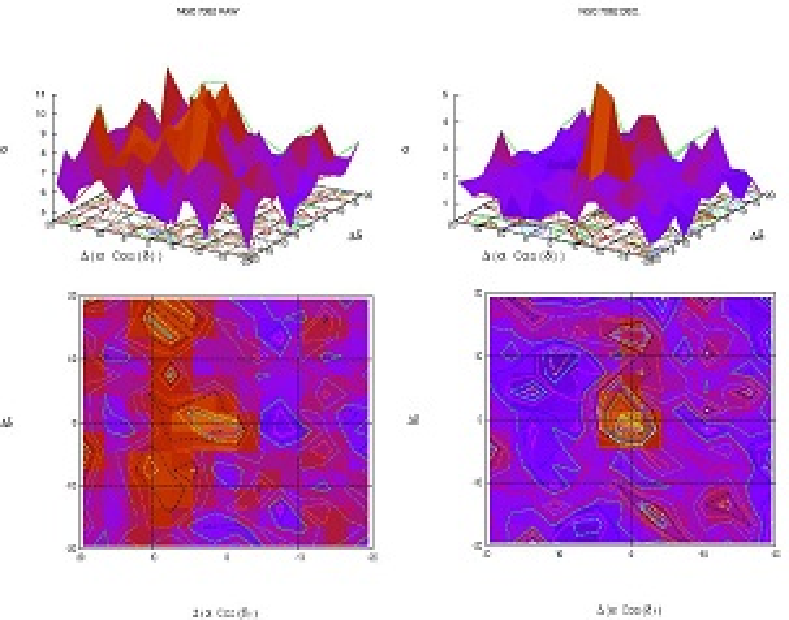}
\caption{For observed (raw) photometry, top panels: stellar surface$-$density 
$\sigma (stars\,\rm arcmin^{-2}$) of \astrobj{NGC~7062}, computed for a mesh size of 
$3^\prime\times3^\prime$, centred on the coordinates in Table 1. 
Bottom panels : The corresponding isopleth surfaces. 
Left panel: observed (raw) photometry. Right panel: decontaminated photometry.}
\end{figure}

\begin{figure}
\centering
\includegraphics*[width = 10cm, height = 10cm]{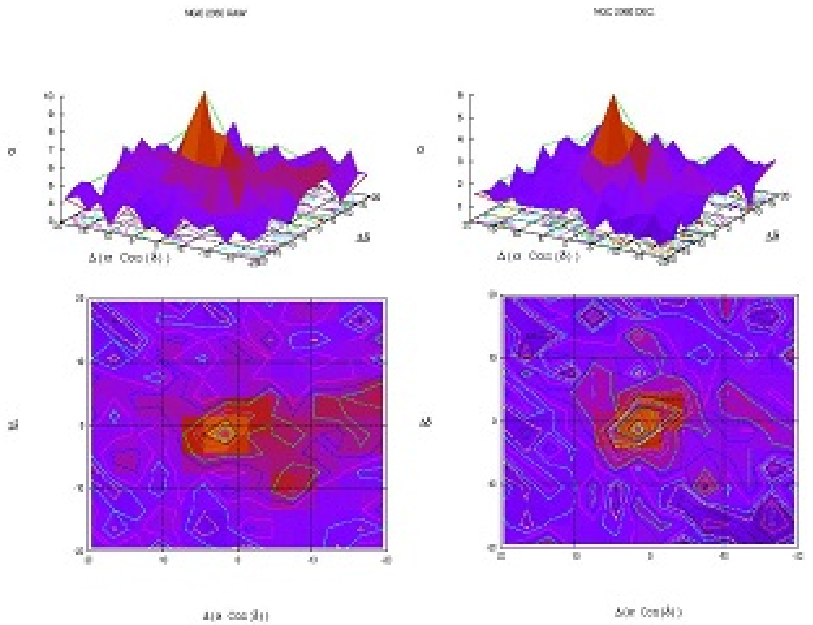}
\caption{For observed (raw) photometry, top panels: stellar surface$-$density 
$\sigma (stars\,\rm arcmin^{-2}$) of \astrobj{NGC~2360}, computed for a mesh size of 
$3^\prime\times3^\prime$, centred on the coordinates in Table 1. 
Bottom panels : The corresponding isopleth surfaces. 
Left panel: observed (raw) photometry. Right panel: decontaminated photometry.}
\end{figure}

\begin{figure}
\centering
\includegraphics*[width = 4cm, height = 7cm]{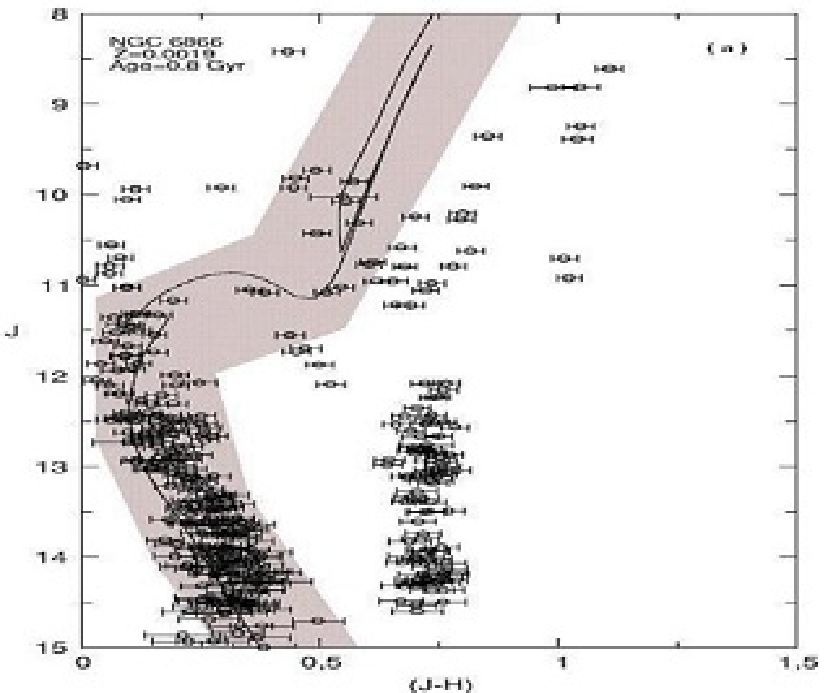}
\includegraphics*[width = 4cm, height = 7cm]{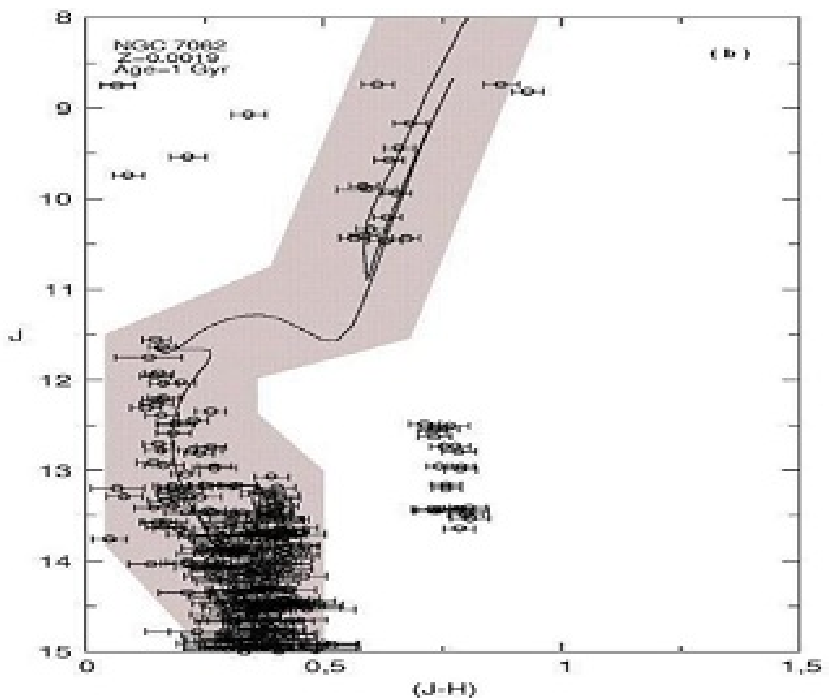}
\includegraphics*[width = 4cm, height = 7cm]{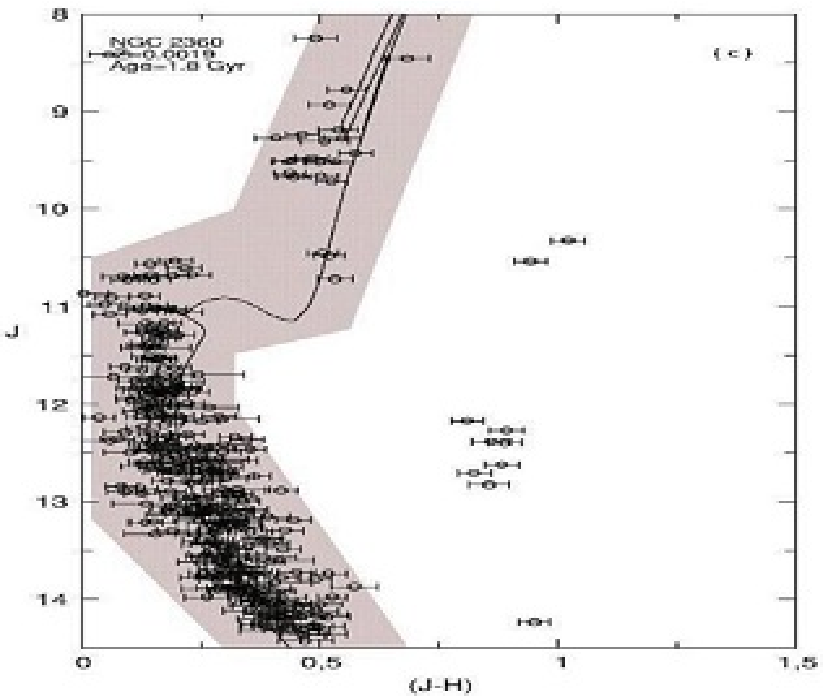}
\caption{Observed decontaminated $J\times(J-H)$ CMDs  extracted from the regions of 
$R=19'.07$ for \astrobj{NGC~6866}, $R=7'.53$ for \astrobj{NGC~7062} and $R=22'.63$ for \astrobj{NGC~2360}, respectively.
The solid blue lines in the panels represent the fitted 0.8, 1, 1.8 Gyr  Padova isochrones 
for $Z= +0.019$ (solar) abundance. The CMD filter used to isolate cluster 
MS/evolved stars is shown with the shaded area.}
\end{figure}

\begin{figure}
\centering
\includegraphics*[width = 4cm, height = 7cm]{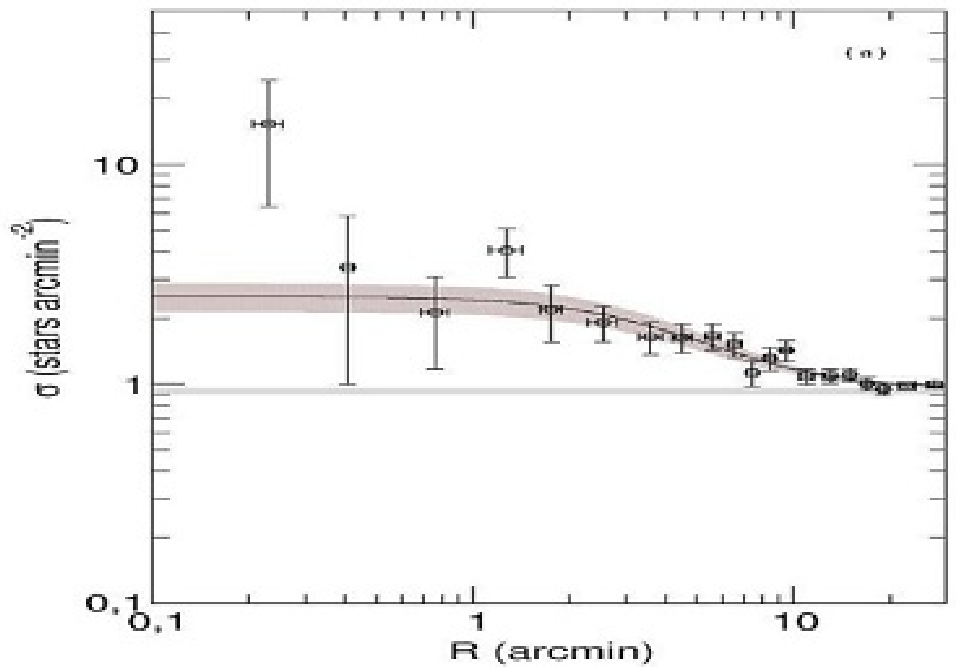}
\includegraphics*[width = 4cm, height = 7cm]{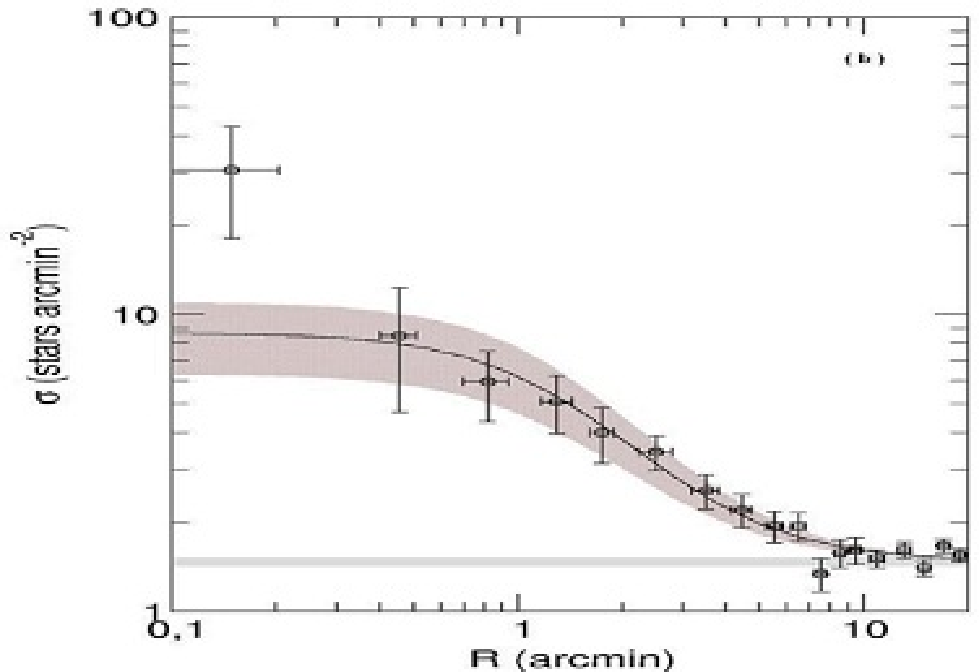}
\includegraphics*[width = 4cm, height = 7cm]{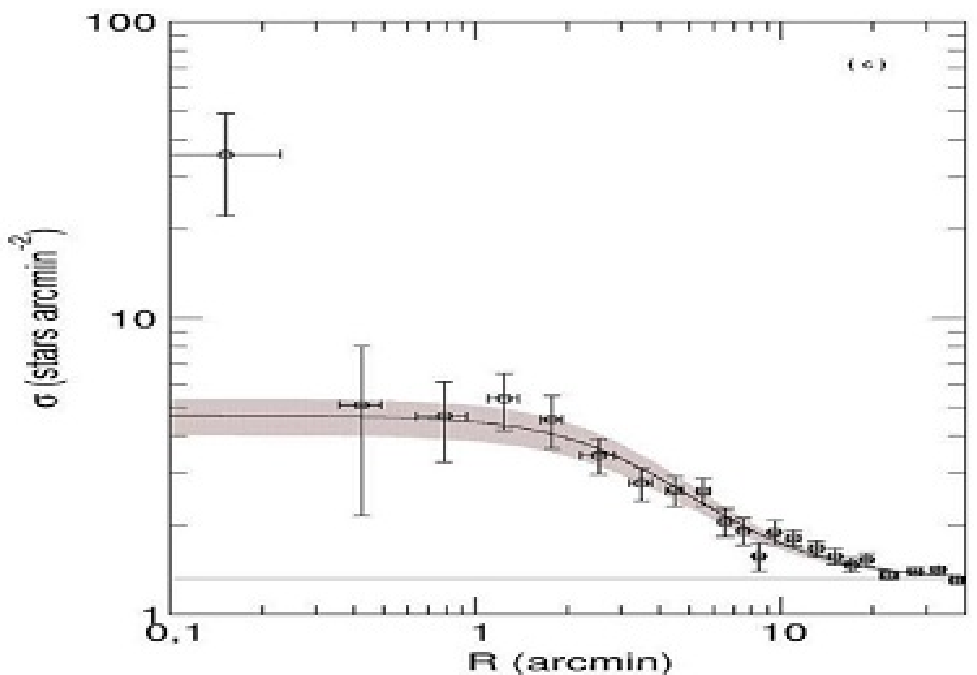}
\caption{Stellar RDPs (open circles) of \astrobj{NGC~6866}, \astrobj{NGC~7062}, \astrobj{NGC~2360} built with CMD filtered photometry. 
Solid line shows the best-fit King profile. Horizontal red bar: stellar background level measured 
in the comparison field. Shaded region: $1\sigma$ King fit uncertainty.}
\end{figure}

\begin{figure}
\centering
\includegraphics*[width = 8cm, height = 11cm]{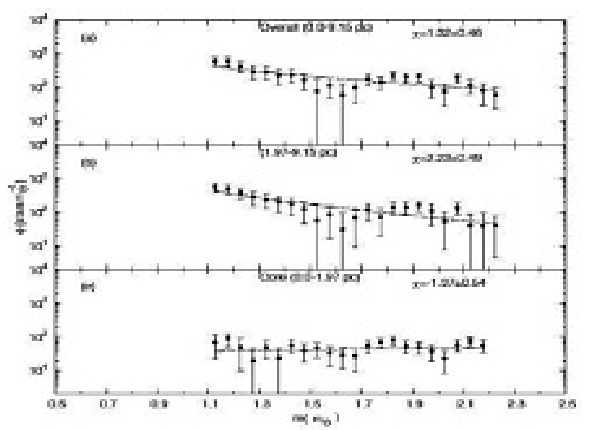}
\caption{$\phi(m)(stars ~m_\odot^{-1})$ versus $m_\odot$ of \astrobj{NGC~6866} cluster,
as a function of distance from the core.}
\end{figure}

\begin{figure}
\centering
\includegraphics*[width = 8cm, height = 11cm]{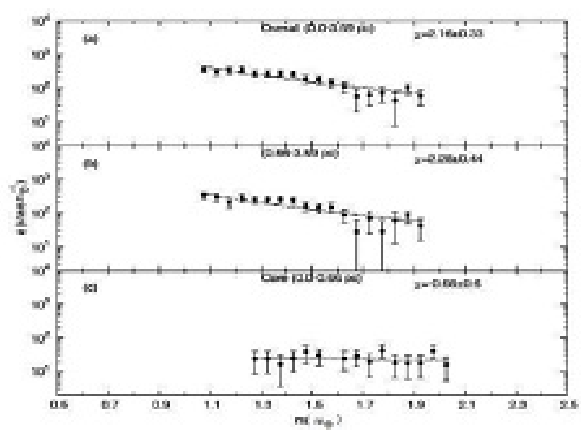}
\caption{$\phi(m)(stars ~m_\odot^{-1})$ versus $m_\odot$ of \astrobj{NGC~7062} cluster,
as a function of distance from the core.}
\end{figure}

\begin{figure}
\centering
\includegraphics*[width = 8cm, height = 11cm]{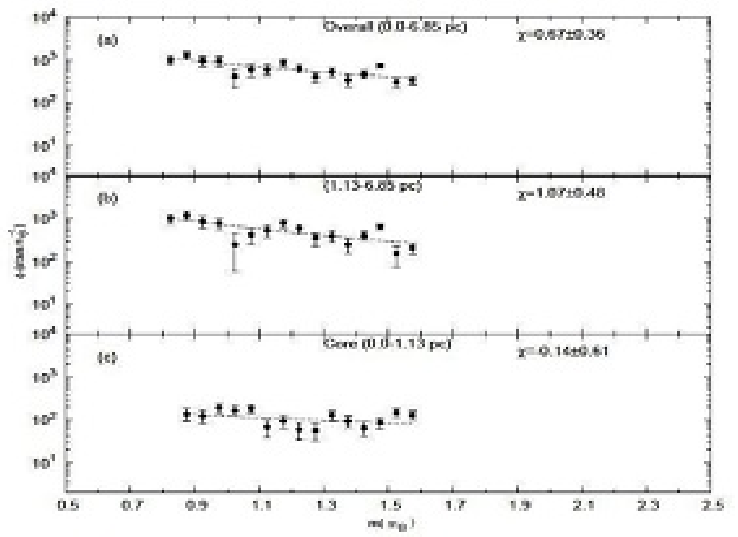}
\caption{$\phi(m)(stars ~m_\odot^{-1})$ versus $m_\odot$ of \astrobj{NGC~2360} cluster,
as a function of distance from the core.}
\end{figure}

\begin{figure}
\centering
\includegraphics*[width = 8.5cm, height = 8.5cm]{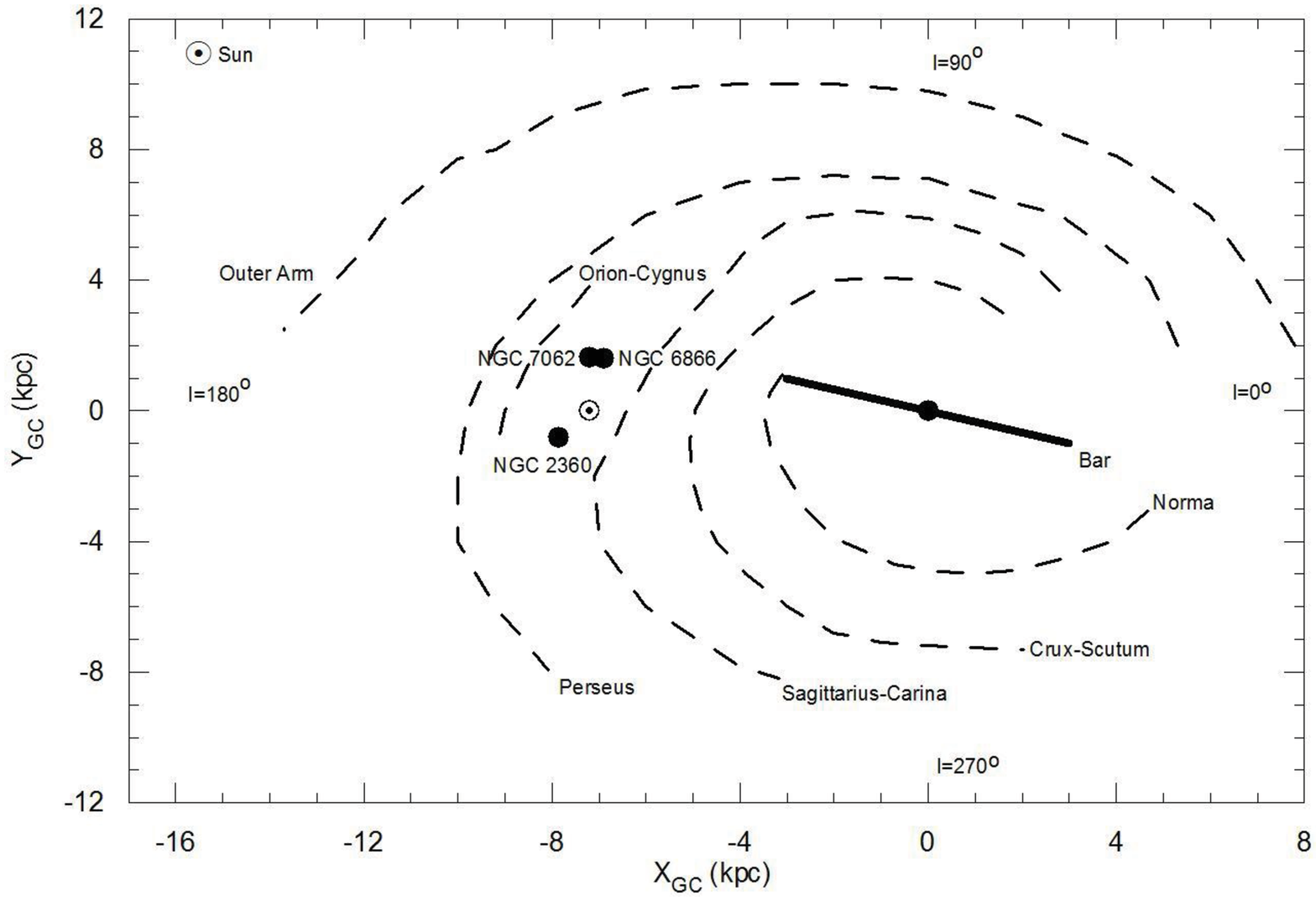}
\caption{Spatial distribution of \astrobj{NGC~6866}, \astrobj{NGC~7062}, \astrobj{NGC~2360} clusters (filled circles). 
The schematic projection of the Galaxy is seen 
from the North pole. The Sun's distance to the 
Galactic center is taken to be 7.2 kpc of \cite{bic06b}.}
\end{figure}

\begin{figure}
\centering
\includegraphics*[width = 12cm, height = 12cm]{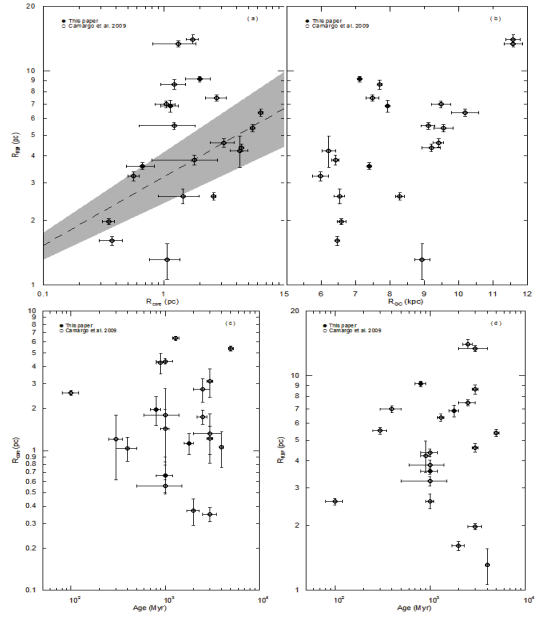}
\caption{$R_{RDP}-R_{core}$, $R_{RDP}-R_{GC}$, $R_{core}-Age~(Myr)$, and  $R_{RDP}- Age~(Myr)$ relations of 
\astrobj{NGC~6866}, \astrobj{NGC~7062}, \astrobj{NGC~2360} clusters. 
Filled and empty circles show the clusters of this work and \cite{cam09}, respectively.}
\end{figure}

\clearpage

\begin{table*}
\tiny
{\footnotesize
\caption{Literature and presently optimised coordinates.}
\label{tab1}
\tiny
\renewcommand{\tabcolsep}{3.6mm}
\renewcommand{\arraystretch}{1.1}
\begin{tabular}{lrrrrrrrrr}
\hline
\hline
\multicolumn{5}{c}{Literature}&\multicolumn{5}{c}{This paper}\\
\cline{2-5}
\cline{7-10}
\tiny
Cluster&$\alpha(2000)$&$\delta(2000)$&$\ell$&$b$&&$\alpha(2000)$&$\delta(2000)$&$\ell$&$b$\\
&(h\,m\,s)&$(^{\circ}\,^{\prime}\,^{\prime\prime})$&$(^{\circ})$&$(^{\circ})$&&(h\,m\,s)&$(^{\circ}\,^{\prime}\,^{\prime\prime})$&$(^{\circ})$&$(^{\circ})$ \\
\hline
NGC\,6866 &20 03 55&+44 09 30&79.56&6.84& &20 04 00.58&+44 09 13.50&79.56&6.82\\
NGC\,7062 &21 23 27&+46 22 42&89.96&-2.75& &21 23 22.64&+46 23 48.76&89.96&-2.73\\
NGC\,2360 &07 17 43&-15 38 30&229.81&-1.43& &07 17 42.95&-15 37 55.50&229.80&-1.42\\
\hline
\end{tabular}
}
\end{table*}

\begin{table}
{\scriptsize
\caption{Derived fundamental astrophysical parameters, Age~(Gyr), $E(B-V)$, $(V-Mv)_{o}$, $d$ (kpc) from 
2~MASS JH${K_{s}}$ photometry and  the isochrones of $Z_{\odot}$. 
$R_{GC}$~(kpc) in Col.~6 is the Galactocentric distance of three clusters.}
\renewcommand{\tabcolsep}{1.0mm}
\renewcommand{\arraystretch}{0.5}
\begin{tabular}{lccccccc}
\hline
Cluster&Age~(Gyr)&$E(B-V)$&$(V-Mv)_{o}$&$d$~(kpc)&$R_{GC}$~(kpc)\\
($1$)&($2$)&($3$)&($4$)&($5$)&($6$)\\
\hline
NGC\,6866 &$0.8\pm0.1$&$0.19\pm0.06$&$11.08\pm0.11$&$1.65\pm0.09$&$7.11\pm0.02$\\
NGC\,7062 &$1.0\pm0.2$&$0.32\pm0.06$&$11.07\pm0.11$&$1.64\pm0.09$&$7.40\pm0.02$\\
NGC\,2360 &$1.8\pm0.2$&$0.06\pm0.03$&$10.09\pm0.06$&$1.04\pm0.03$&$7.93\pm0.02$\\
\hline
\end{tabular}
}
\end{table}

\begin{table}
\tiny
\centering
\tiny
{\footnotesize
\tiny
\caption{Comparison of the fundamental astrophysical parameters of 
\astrobj{NGC~6866}, \astrobj{NGC~7062}, \astrobj{NGC~2360} to the literature. Cols.~1$-$4 
represent the cluster name, reddening, the metal and heavy element 
abundances, respectively.  True distance modulus values, $(V-Mv)_{o}$, 
and their corresponding heliocentric distances 
are presented in Cols.~5$-$6, respectively.  Cols.~7$-$8 give the
age of $\log(A)$ and A~(Gyr), respectively. The isochrones/observational 
ZAMSs and the photometry of the literature are
listed in Cols.~9$-$10, respectively. The references of the literature are given in Col.~11.}
\tiny
\begin{tabular}{ccccccccccc}
\hline
 Cluster &E(B-V) &[Fe/H] & Z & $(V-Mv)_{o}$ &$d$~(kpc) & $\log(A)$ & $A~(Gyr)$ &$Iso/obs$ ZAMS & Photometry &  Reference \\
($1$)&($2$)&($3$)&($4$)&($5$)&($6$)&($7$)&($8$)&($9$)&($10$)&($11$)\\
\hline
\multicolumn{ 1}{c}{NGC 6866} & 0.19 &  -    &0.019&11.08&1.65&8.90 &0.80  &G02            &2~MASS & This paper \\
& & & & & & & & & & \\
\multicolumn{ 1}{c}{}         & 0.17 &  -    &0.019&10.80&1.45&8.58 &0.38  &obs.ZAMS       & UBV  &      [1] \\
\hline
\multicolumn{ 1}{c}{NGC 7062} & 0.32 &  -    &0.019&11.07&1.64&9.00 &1.00  &G02            &2~MASS & This paper \\
& & & & & & & & & & \\
\multicolumn{ 1}{c}{}         &0.46  &$-$0.35&0.003&12.18&2.73&8.45 &0.28  &Vandenberg~1985&$uvby-\beta$ & [2] \\
\hline
\multicolumn{ 1}{c}{NGC 2360} &0.06 &   -    &0.019&10.09&1.04&9.25 &1.80  &G02           &2~MASS    & This paper \\
& & & & & & & & & & \\
\multicolumn{ 1}{c}{}         &0.07  &  -     &0.019&10.30&1.15& -   &-     &Eggen ZAMS     & UBV   & [3] \\
\hline
\end{tabular}
\\
$[1]$: \cite{hoa61}, $[2]$: \cite{pen90}, 
$[3]$: \cite{egg68}\\}
\end{table}

\begin{table*}
{\footnotesize
\tiny
\begin{center}
\caption{Structural parameters of the clusters of \astrobj{NGC~6866}, \astrobj{NGC~7062}, \astrobj{NGC~2360}.}
\tiny
\renewcommand{\tabcolsep}{1.1mm}
\renewcommand{\arraystretch}{1.1}
\begin{tabular}{llllllllllll}
\hline
\tiny
Cluster&$(1')$&$\sigma_{0K}$&$\sigma_{bg}$&$R_{core}$&$R_{RDP}$&$\sigma_{0K}$&$\sigma_{bg}$&$R_{core}$&$R_{RDP}$&${\Delta}R$&CC\\
&($pc$)&($*\,pc^{-2}$)&($*\,pc^{-2}$)&($pc$)&($pc$)&($*\,\prime^{-2}$)&($*\,\prime^{-2}$)&($\prime$)&($\prime$)&($\prime$)&\\
($1$)&($2$)&($3$)&($4$)&($5$)&($6$)&($7$)&($8$)&($9$)&($10$)&($11$)&($12$)\\
\hline
NGC\,6866&$0.48$&$6.83\pm1.65$&$4.08\pm0.12$&$1.97\pm0.46$&$9.15\pm0.27$&$1.52\pm0.37$&$0.94\pm0.02$&$4.13\pm0.94$&$19.07\pm0.57$&$50-60$&0.90\\
NGC\,7062&$0.48$&$31.61\pm10.37$&$6.45\pm0.20$&$0.66\pm0.17$&$3.59\pm0.13$&$7.20\pm2.36$&$1.47\pm0.05$&$1.38\pm0.35$&$7.53\pm0.27$&$25-35$&0.89\\
NGC\,2360&$0.30$&$36.89\pm7.27$&$14.45\pm0.26$&$1.13\pm0.19$&$6.85\pm0.44$&$3.37\pm0.66$&$1.32\pm0.02$&$3.73\pm0.63$&$22.63\pm1.44$&$40-50$&0.93\\
\hline
\end{tabular}
\begin{list}{Table Notes.}
\item Col. 2: arcmin to parsec scale. To minimize degrees of freedom in RDP fits with the King-like profile (see text), $\sigma_{bg}$ was kept fixed (measured in the respective comparison fields) while $\sigma_{0}$ and $R_{core}$ were allowed to vary. Col. 11: comparison field ring. Col. 12: correlation coefficient.
\end{list}
\end{center}
}
\end{table*}

\begin{table*}
\tiny
{\footnotesize
\caption{The mass information for the clusters of \astrobj{NGC~6866}, \astrobj{NGC~7062}, \astrobj{NGC~2360}.}
\label{tab1}
\tiny
\renewcommand{\tabcolsep}{0.6mm}
\renewcommand{\arraystretch}{1.1}
\begin{tabular}{cccccccccc}
\hline
\hline
\multicolumn{3}{c}{Evolved}&\multicolumn{1}{c}{}&\multicolumn{2}{c}{Observed+Evolved}&\multicolumn{2}{c}{Extrapolated+Evolved}\\
\cline{2-3}
\cline{4-5}
\cline{5-8}
\tiny
Region&$N^{*}$&$m_{evol}$&$\chi$  &$N^{*}$&$m_{obs}$&$N^{*}$&$m_{tot}$\\
(pc)&(Stars)&($10^1 M_{\odot}$)&&($10^2 Stars$)&($10^2 M_{\odot}$)&($10^2 Stars$)&($10^2 M_{\odot}$)\\
(1)&(2)&(3)&(4)&(5)&(6)&(7)&(8)\\
\hline
&NGC~6866&&$m=1.13-2.23~m_{\odot}$&&&&\\
\hline
0.00 - 1.97&$6\pm3$&$1.3\pm0.6$&$-1.37\pm0.54$&$0.65\pm0.07$&$1.12\pm0.40$&$1.04\pm0.50$&$1.38\pm0.44$\\
1.97 - 9.15&$7\pm8$&$1.5\pm1.7$&$2.23\pm0.49$&$2.01\pm0.24$&$3.02\pm0.87$ &$48.8\pm37.5$&$16.9\pm7.15$\\
0.00 - 9.15&$13\pm8$&$2.8\pm1.9$&$1.32\pm0.46$&$2.59\pm0.25$&$4.05\pm1.13$&$48.8\pm37.0$&$17.8\pm7.08$\\
\hline
&NGC~7062&&$m=1.08-1.93~m_{\odot}$&&&&\\
\hline
0.00 - 0.66&$1\pm1$&$0.1\pm0.2$&$-0.66\pm0.60$&$0.33\pm0.04$&$0.53\pm0.23$&$1.08\pm2.49$&$0.98\pm0.72$\\
0.66 - 3.59&$10\pm5$&$2.1\pm1$&$2.28\pm0.44$&$1.65\pm0.12$&$2.33\pm0.45$  &$39.7\pm30.2$&$13.5\pm5.66$\\
0.00 - 3.59&$11\pm5$&$2.2\pm1.1$&$2.16\pm0.33$&$1.94\pm0.13$&$2.73\pm0.42$&$45.7\pm34.5$&$15.6\pm6.43$\\
\hline
&NGC~2360&&$m=0.83-1.63~m_{\odot}$&&&&\\
\hline
0.00 - 1.13&$22\pm5$&$3.5\pm0.8$&$-0.14\pm0.61$&$1.15\pm0.09$&$1.46\pm0.21$&$-$&$-$\\
1.13 - 6.85&$19\pm12$&$3.0\pm1.9$&$1.07\pm0.48$&$4.74\pm0.37$&$5.38\pm0.58$&$51.9\pm39.8$&$18.1\pm7.49$\\
0.00 - 6.85&$41\pm13$&$6.5\pm2.1$&$0.67\pm0.36$&$5.81\pm0.37$&$6.77\pm0.60$&$47.1\pm33.5$&$18.1\pm6.24$\\
\hline
\end{tabular}
\\
Col.~1: the distance from the core. Cols.~2,5,7 : cluster stars for the regions in Col.~1.
Col.~4 gives the MF slopes ($\chi$),  derived for the low-mass and high-mass ranges. 
The masses of $m_{evol}$, $m_{obs}$, and $m_{tot}$ are listed in Cols.~3, 6 and 8, respectively.
\\
}
\end{table*}

\begin{table*}
\centering
\tiny
{\footnotesize
\caption{Relaxation time and evolutionary parameter}
\label{tab1}
\tiny
\renewcommand{\tabcolsep}{0.6mm}
\renewcommand{\arraystretch}{1.1}
\begin{tabular}{ccccc}
\hline
\hline
\multicolumn{3}{c}{Core}&\multicolumn{2}{c}{Overall}\\
\cline{2-3}
\cline{4-5}
\tiny
Cluster&$t_{relax}$&$\tau$&$t_{relax}$&$\tau$\\
&(Myr)&&$(Myr)$&\\
(1)&(2)&(3)&(4)&(5)\\
\hline
NGC 6866&$1.77\pm0.78$&$453\pm207$&$210\pm141$&$3.80\pm2.59$\\
NGC 7062&$0.61\pm1.12$&$1640\pm3020$&$77.8\pm52.0$&$12.8\pm8.96$\\
NGC 2360&$3.17\pm71.7$&$569\pm12900$&$153\pm96.1$&$11.8\pm7.55$\\
\hline
\end{tabular}
}
\end{table*}

\section{Acknowledgments}
We thank the anonymous referee for her/his comments and suggestions. 
This publication makes use of data products from the Two Micron All Sky Survey, 
which is a joint project of the University of Massachusetts and the 
Infrared Processing and Analysis Centre/California Institute of Technology, 
funded by the National Aeronautics and Space Administration and the National 
Science Foundation. This research has made use of the WEBDA database, operated 
at the Institute for Astronomy of the University of Vienna. 
This work was supported by the Research Fund of the University of Istanbul,
Project number: BYP-22411.

\clearpage

\end{document}